\documentclass[12pt]{article} 
\usepackage{amssymb,amsmath}
\usepackage{epsf}
\usepackage{graphicx} 

\begin{document}

\title{Energy Relaxation in Fermi-Pasta-Ulam Arrays}

\author{R. Reigada\\
Departament de Qu\'{\i}mica F\'{\i}sica\\
Universitat de Barcelona\\
Avda. Diagonal 647, 08028 Barcelona, Spain\\
\and
A. Sarmiento\\
Instituto de Matem\'aticas\\
Universidad Nacional Aut\'onoma de M\'exico\\
Ave. Universidad s/n, 62200 Chamilpa Morelos, M\'exico\\
\and
Katja Lindenberg\\
Department of Chemistry and Biochemistry and\\
Institute for Nonlinear Science\\
University of California, San Diego, La Jolla, California 92093}

\date{\today}

\maketitle

\begin{abstract}
The dynamics of energy relaxation in thermalized one- and two-dimensional
arrays with nonlinear interactions depend in detail on the interactions
and, in some cases, on dimensionality.  We describe and explain
these differences for arrays of the Fermi-Pasta-Ulam type. In
particular, we focus on the roles of harmonic contributions to the
interactions and of breathers in the relaxation process.  
\end{abstract} 

\bigskip

\noindent
PACS number(s): 05.40.-a, 05.45.-a, 63.20.Pw
\bigskip

\section{Introduction}

The ability of extended systems to support
the localization and transport of vibrational
energy has been invoked in a number of physical settings including
DNA molecules~\cite{peyrard}, hydrocarbon structures~\cite{kopidakis},
energy storage and transport in proteins~\cite{pro1,pro2,pro3}, the
creation of vibrational intrinsic localized modes in anharmonic crystals
using an optimally chosen sequence of femtosecond laser
pulses~\cite{rossler}, photonic crystal waveguides~\cite{mingaleev}, and
targeted energy transfer between donors and acceptors in
biomolecules~\cite{aubry}.
It has become increasingly clear that thermal fluctuations may strongly
affect (sometimes leading to degradation but at times actually
helping) the process of energy localization and energy
mobility~\cite{peyrard,kopidakis,bourbonnais,piazza,harvesting,cretegny,wepulse,wethermalresonance,burlakov,kosevich,floria}.
It is thus clearly important to investigate the nature and dynamics
of thermal fluctuations in nonlinear arrays.  The understanding of
the spatial and temporal
evolution of the thermal relaxation landscape will in turn lead to a better
understanding of other chemical and physical processes that may be
occurring in the relaxing landscape.

The spatial and temporal evolution of an influx of energy into or
an efflux of energy 
out of a nonlinear array, and the dynamical pathways that characterize 
energy relaxation in such an array, depend on a large number of
factors (for recent reviews of a huge literature see~\cite{FW,bri9}).
The nature of the interactions and where the nonlinearities reside
(in the local potentials or in the interactions), the boundary
conditions (free, fixed, or periodic), the size of the system, its
dimensionality, the way in which energy is deposited in the system (initial
conditions), etc., can all influence the evolution profoundly, and no
general formalism that encompasses all these variations has yet been
developed.  It is thus necessary in this study (as in most others) to
circumscribe the range of our inquiry.  We have been
particularly interested in discrete extended systems in
which localized energy can also be {\em
mobile}~\cite{harvesting,wepulse,wethermalresonance}, and so our studies
have focused on arrays with hard nonlinear interactions and no
local potentials, specifically on Fermi-Pasta-Ulam (FPU) lattices. 
Some movement of localized energy may also occur in arrays with soft local 
potentials (see e.g.~\cite{peyrard}), but it is much more difficult to
achieve.  We thus concentrate on FPU arrays.

Specifically in this paper we study energy relaxation in
one-dimensional~\cite{new}
and two-dimensional FPU arrays with quartic potentials.  The Hamiltonian
in one dimension is
\begin{equation}
H = \sum_{i=1}^{N} \frac{\dot{x}_{i}^2}{2} + \frac{k}{2} \sum_{i=1}^{N}
(x_{i} - x_{i-1})^2 +\frac{k^\prime}{4} \sum_{i=1}^{N} (x_{i} - x_{i-1})^4
\label{ham1}
\end{equation}
where $N$ is the number of sites;  $k$ and $k^\prime$ are the harmonic
and anharmonic force constants, respectively. In two dimensions for an
$N\times N$ lattice
\begin{eqnarray}
H = \sum_{i,j=1}^{N} \frac{\dot{x}_{i,j}^2}{2}
&&+ \frac{k}{2} \sum_{i,j=1}^{N}
\left[(x_{i,j} - x_{i-1,j})^2 + (x_{i,j} - x_{i,j-1})^2\right]
\nonumber\\
&&+\frac{k^\prime}{4} \sum_{i,j=1}^{N}
\left[(x_{i,j} - x_{i-1,j})^4 +(x_{i,j} - x_{i,j-1})^4\right].
\label{ham2}
\end{eqnarray}
To study energy relaxation we
initially thermalize the system at temperature
$T$ (see below), then connect the boundary sites (two end sites for a
1d system, $4(N-1)$ edge sites for 2d arrays) to a zero-temperature
reservoir via damping terms, and observe the thermal relaxation of the array
toward zero temperature~\cite{piazza,new,tsi1,tsi2}.  We find that 
relaxation occurs through an interesting cascade of decay times that is
sensitively dependent on the precise form of the interactions.  This leads
us to focus on two issues in the relaxation process: 1) the explicit role
of the harmonic terms vs. anharmonic terms in the FPU Hamiltonian,
and 2) the effects of array dimensionality.  

To thermalize the system to a given temperature $T$ we augment the
equations of motion resulting from Eqs.~(\ref{ham1}) with the Langevin
prescription connecting each site to a heat bath.  In one dimension
\begin{equation}
\ddot{x_i} = -\frac{\partial}{\partial x_i} [V(x_i -x_{i-1}) +
V(x_{i+1}-x_i)] -\gamma_0 \dot{x}_i +\eta_i(t).
\label{langfinitet}
\end{equation}
Here $V(x_i-x_j)$ is the FPU potential, $\gamma_0$ is the
dissipation parameter, and the $\eta_i(t)$ are mutually uncorrelated
zero-centered Gaussian
$\delta$-correlated fluctuations that satisfy the fluctuation-dissipation
relation at temperature $T$:
\begin{equation}
\langle \eta_i(t) \rangle = 0, \;\;\;\;\;\;\;\;\;\;\;\;
\langle \eta_i(t) \eta_j(t') \rangle = 2 \gamma_0 k_B T
\delta_{ij}\delta(t-t').
\label{fdr}
\end{equation}
The brackets here and below denote an ensemble average, and
$k_B$ is Boltzmann's constant.  The generalization to two dimensions is
immediate.  We implement free-end boundary conditions, that is,
$x_0=x_1$ and $x_N=x_{N+1}$ in one dimension and $x_{0,j}=x_{1,j}$,
$x_{N,j}=x_{N+1,j}$, $x_{i,0}=x_{i,1}$, and $x_{i,N}=x_{i,N+1}$
in two dimensions. For the integrations of the equations of motion
we use the fourth order Runge-Kutta method.

The thermalization process involves the spontaneous emergence of anharmonic
(including localized) modes.  In order to understand the dynamics of such
modes, we begin again with a thermalized array and explicitly inject 
a breather-like excitation of energy much higher than the thermal energy.
We then observe how the entire system, thermalized array plus injected
excitation, evolves and relaxes toward zero temperature.

In Sec.~\ref{laytheground} we itemize the measures used to display the
relaxation process.  In Sec.~\ref{results} we present our relaxation
results, obtained primarily from numerical simulations. 
Section~\ref{breather} shows what happens to an explicitly injected local
excitation as the arrays relax to zero temperature.  Section~\ref{summary}
provides a brief summary and synthesis of the outcomes.

\section{Measures of Thermal Relaxation}
\label{laytheground}

Once our system is thermalized to temperature $T$ we disconnect it from the
thermal bath (i.e., we remove the
$\eta_i(t)$ and $\gamma_0\dot{x}_i$ terms from Eq.~(\ref{langfinitet}) or
the equivalent terms from the two-dimensional set of equations),
and we connect the edge sites to a cold $T=0$ reservoir through
the addition of dissipative terms $-\gamma\dot{x}$ 
to the equations of motion for these sites.  We then continue
the integration using the thermalized positions and displacements as the
initial conditions.  An ensemble is constructed by repeating this
experiment for different thermalized initial conditions.
There are of course a variety of ways to display the
outcome, and in our work we have chosen three.  One, the most ``global"
measure, is to follow the decay of the total array energy $E(t)$ as
a function of time.  The second is to follow the spectrum of the system as
a function of time.  This gives a frequency-by-frequency picture of the
relaxation process and therefore a more complete description.  The third is
to show a spatial energy landscape as a function of time. 

The total energy $E(t)$ of the array is simply the Hamiltonian function
evaluated with the displacements and velocities obtained from the
equations of motion.  For a one-dimensional {\em harmonic}
array this function has been
calculated analytically by Piazza et al.~\cite{piazza}
for small $\gamma$ and large $N$: 
\begin{align}
\frac{E(t)}{E(0)} & = \frac{1}{\pi} \int_0^\pi dq~e^{-2t/\tau(q)}
=e^{-t/{\tau_0}}I_0(t/\tau_0)\nonumber\\ \nonumber\\
&=
\begin{cases}
e^{-t/\tau_0} &\text{for $t \ll \tau_0$}, \\
\left(\frac{\displaystyle 2\pi t}{\displaystyle
\tau_0}\right)^{-1/2} &\text{for $t \gg \tau_0$}.
\end{cases}
\label{energy1}
\end{align}
In the first line $I_0$ is the modified zero-order Bessel function, and
$\tau_0=N/2\gamma$.  The decay time $\tau(q)$ for phonons of wavevector $q$
is
\begin{equation}
\frac{1}{\tau(q)}=\frac{1}{\tau_0} \cos^2 \left(\frac{q}{2}\right).
\end{equation}
The second line in Eq.~(\ref{energy1}) gives the short-time and long-time
behaviors.  The former is a simple exponential decay associated with the
lowest frequency phonon mode since it has the shortest decay time.
The power law relaxation arises from the cascade of different
decay times of the different phonon modes.  For finite $N$ 
the decay becomes exponential again when only the modes near the
band-edge of the spectrum still survive.  Note that the decay is not
exponential throughout
(a common misapprehension for harmonic systems), although the initial
exponential behavior does last longer the larger the system.
Each phonon mode decays separately (exponentially) and independently of
all the others.  This translates to
an independent and separate decay of each frequency portion of the spectrum
(see below).  The calculation of the total energy is more cumbersome but
basically similar for two-dimensional arrays.  One finds
that the decay time for phonons of wavevector $\mathbf{q} = (q_x, q_y)$ is
\begin{equation}
\frac{1}{\tau(\mathbf{q})}=\frac{1}{\tau_0} \left[\cos^2
\left(\frac{q_x}{2}\right) + \cos^2 \left(\frac{q_y}{2}\right) \right],
\end{equation}
which upon integration over wavevectors leads to
\begin{equation}
\frac{E(t)}{E(0)}   =e^{-2t/{\tau_0}}I_0^2(t/\tau_0) =
\begin{cases}
e^{-2t/\tau_0} &\text{for $t \ll \tau_0$}, \\
\left(\frac{\displaystyle 2\pi t}{\displaystyle
\tau_0}\right)^{-1} &\text{for $t \gg \tau_0$}.
\end{cases}
\label{energy2}
\end{equation}
The behavior of $E(t)/E(0)$ for 1d and 2d {\em anharmonic}
arrays is expected
to be different than Eqs.~(\ref{energy1}) and (\ref{energy2}), respectively. 
These behaviors will be presented in the next section.

Our second measure of thermal relaxation focuses on the decay pathways of
the different spectral regions as the array cools down.  We concentrate on
the time evolution of the Fourier transform of the relative
displacement correlation function.  Relative displacements provide a
particularly sensitive measure of how adjacent masses are moving relative
to one another.  In a thermalized array the
spectrum of interest is defined as
\begin{equation}
S(\omega)=2\int_0^{\infty} d\tau~ C(\tau) \cos \omega \tau,
\label{equilS}
\end{equation}
where in one dimension
\begin{equation}
C(\tau)= \frac{1}{(N-1)}\sum_{i=2}^{N} \left<
\delta_i x_i(t+\tau)\delta_i x_i(t)\right>
\label{equilC}
\end{equation}
and $\delta_i x_i(t)$ is the relative displacement
\begin{equation}
\delta_i x_i(t)\equiv x_i(t)-x_{i-1}(t).
\label{relative}
\end{equation}
In two dimensions
\begin{align}
C(\tau) &= \frac{1}{N(N-1)}\sum_{i=2}^N\sum_{j=1}^{N} \left<
\delta_i x_{i,j}(t+\tau) \delta_i x_{i,j}(t)\right> \nonumber\\ \nonumber\\
&+ \frac{1}{N(N-1)}\sum_{i=1}^N\sum_{j=2}^{N} \left< 
\delta_j x_{i,j}(t+\tau) \delta_j x_{i,j}(t)\right>  
\label{equilC2}
\end{align}
where
\begin{equation}
\delta_ix_{i,j}(t) \equiv x_{i,j}(t)-x_{i-1,j}(t), \quad
\delta_jx_{i,j}(t) \equiv x_{i,j}(t)-x_{i,j-1}(t).
\label{relative2}
\end{equation}

The thermal equilibrium spectrum for harmonic arrays in
one~\cite{wethermalresonance} and two
dimensions can be calculated analytically.  In one dimension with periodic
boundary conditions (for sufficiently long chains the boundary conditions
do not affect the equilibrium spectrum)
\begin{equation}
S(\omega)=\frac{4\gamma_0 k_BT}{N} \sum_{q=0}^{N-1} \frac{1-\cos(2\pi
q/N)}{[r_1^2(q) +\omega^2][r_2^2(q)+\omega^2]}
\label{harmonicspectrum1}
\end{equation}
where
\begin{equation}
r_{1,2}(q) = -\frac{\gamma_0}{2} \pm
\sqrt{\left(\frac{\gamma_0}{2}\right)^2 - 4k\sin^2\left(\frac{\pi
q}{N}\right)}.
\end{equation}
In two dimensions
\begin{equation}
S(\omega)=\frac{4\gamma_0 k_BT}{N^2} \sum_{p,q=0}^{N-1} \frac{2-\cos(2\pi
p/N) -\cos(2\pi q/N)}{[r_1^2(p,q) +\omega^2][r_2^2(p,q)+\omega^2]}
\label{harmonicspectrum2}
\end{equation}
where now
\begin{equation}
r_{1,2}(p,q) = -\frac{\gamma_0}{2} \pm
\sqrt{\left(\frac{\gamma_0}{2}\right)^2 - 4k\left[\sin^2\left(\frac{\pi
p}{N}\right) + \sin^2 \left( \frac{\pi q}{N}\right)\right]}.
\end{equation}
For anharmonic chains these spectra must be obtained numerically.

To monitor the decay of the spectrum when the thermalized arrays
are connected to a cold reservoir we introduce the time-dependent spectra
\begin{equation}
S(\omega,t)\equiv 2\int_0^{\tau_{max}} d\tau~C(\tau,t) \cos \omega \tau
\end{equation}
where $\tau_{max}\equiv 2\pi/\omega_{min}$ and $\omega_{min}$ is chosen
for a desired frequency resolution;
the choice $\omega_{min}=0.0982$, corresponding to $\tau_{max}=64$,
turns out to be numerically convenient.
The time-dependent correlation function is actually an average over
the time interval $t-t_0$ to $t$, where we have
chosen $t_0 = 100$ (short enough for the correlation function
not to change appreciably but long enough for statistical
purposes) and is defined as follows in one dimension:
\begin{equation}
C(\tau,t) =
\frac{1}{(N-1)}\sum_{i=2}^{N} \frac{1}{\Delta t}
\int_0^{\Delta t} d\tau ' ~\left< \delta_ix_i(t-\tau ')
\delta_ix_i(t-\tau ' -
\tau) \right>,
\end{equation}
where $\Delta t \equiv t_0-\tau_{max}$.  The generalization to the
two-dimensional case is obvious.

\section{Thermal Relaxation}
\label{results}

We start by presenting two sets of figures, each associated with
one of the measures for relaxation mentioned in the previous
section.  Each set presents both 1d and 2d results.  Along with these
results we present some auxiliary figures that help in the interpretation
of the outcomes.

\begin{figure}[htb]
\begin{center}
\leavevmode
\epsfxsize = 4.in
\epsffile{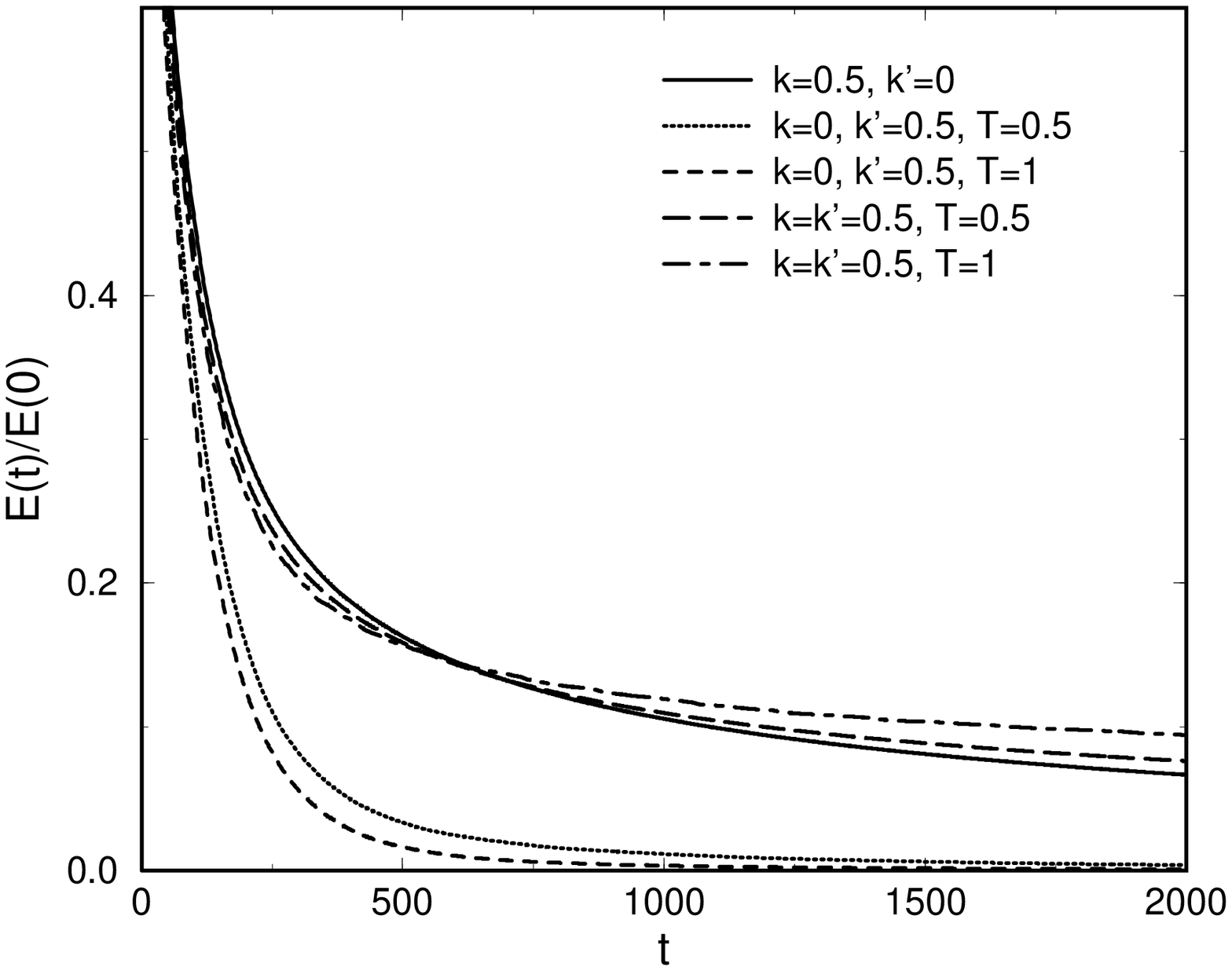}
\leavevmode
\epsfxsize = 4.in
\epsffile{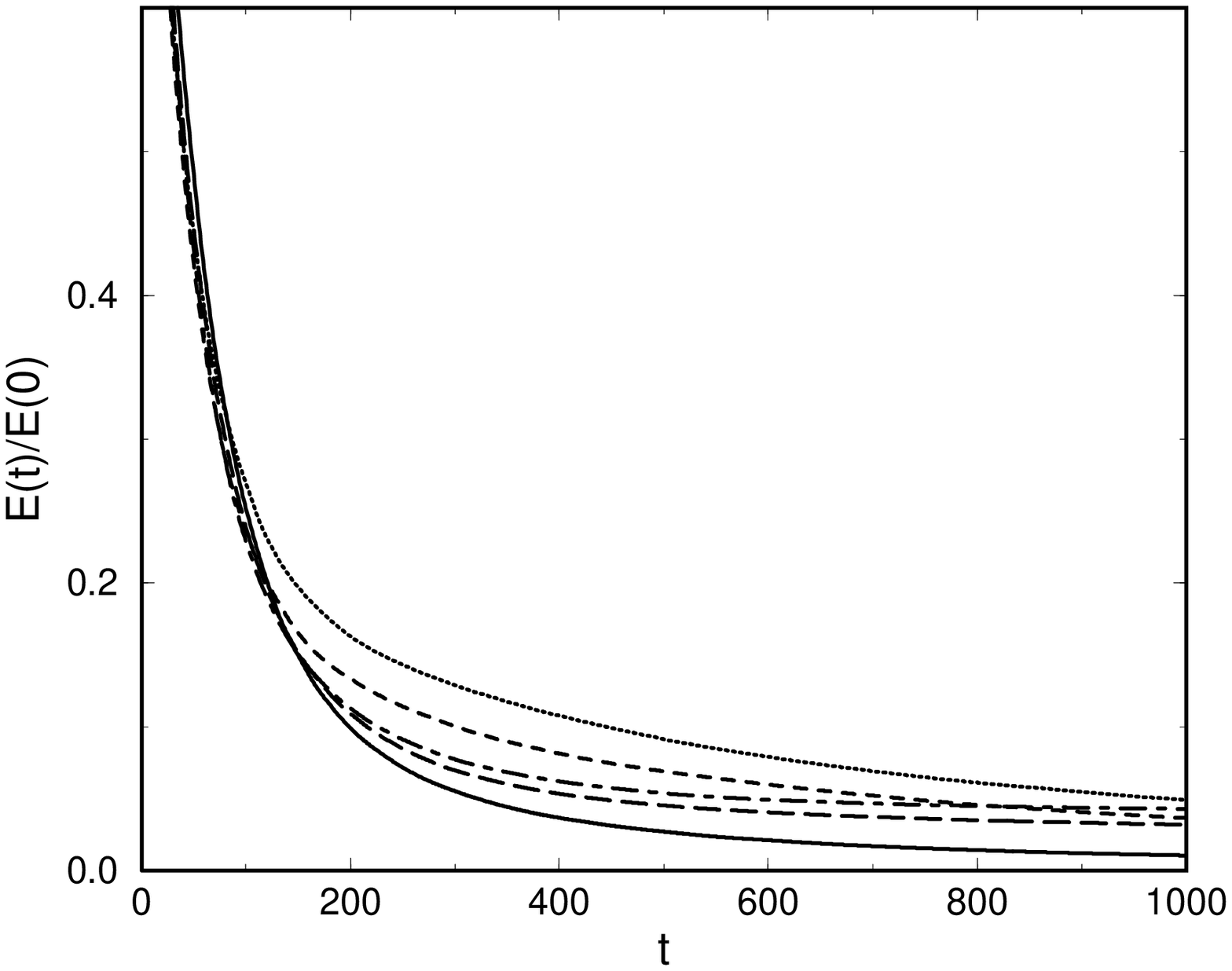}
\end{center}
\caption{Temporal relaxation of the fractional energy for different arrays
with potential parameters and initial temperatures indicated in the panels.
First panel: one dimension ($N=50$).  Second panel: two dimensions
($20\times 20$ lattices).  In all cases $\gamma=0.1$. The fractional energy
for the harmonic arrays is independent of temperature.}
\label{energydecay}
\end{figure}

\begin{figure}[htb]
\begin{center}
\leavevmode
\epsfxsize = 4.in
\epsffile{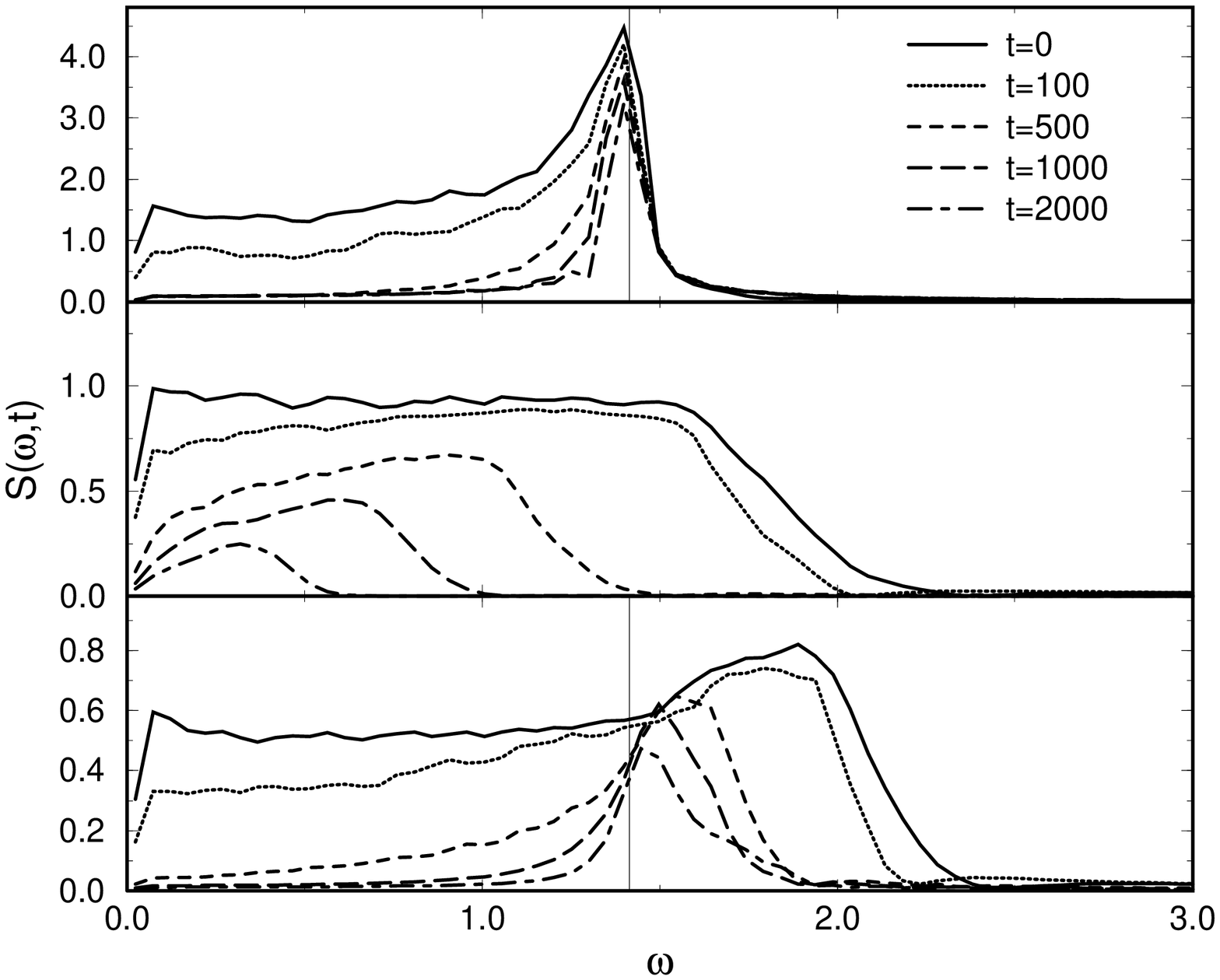}
\leavevmode
\epsfxsize = 4.in
\epsffile{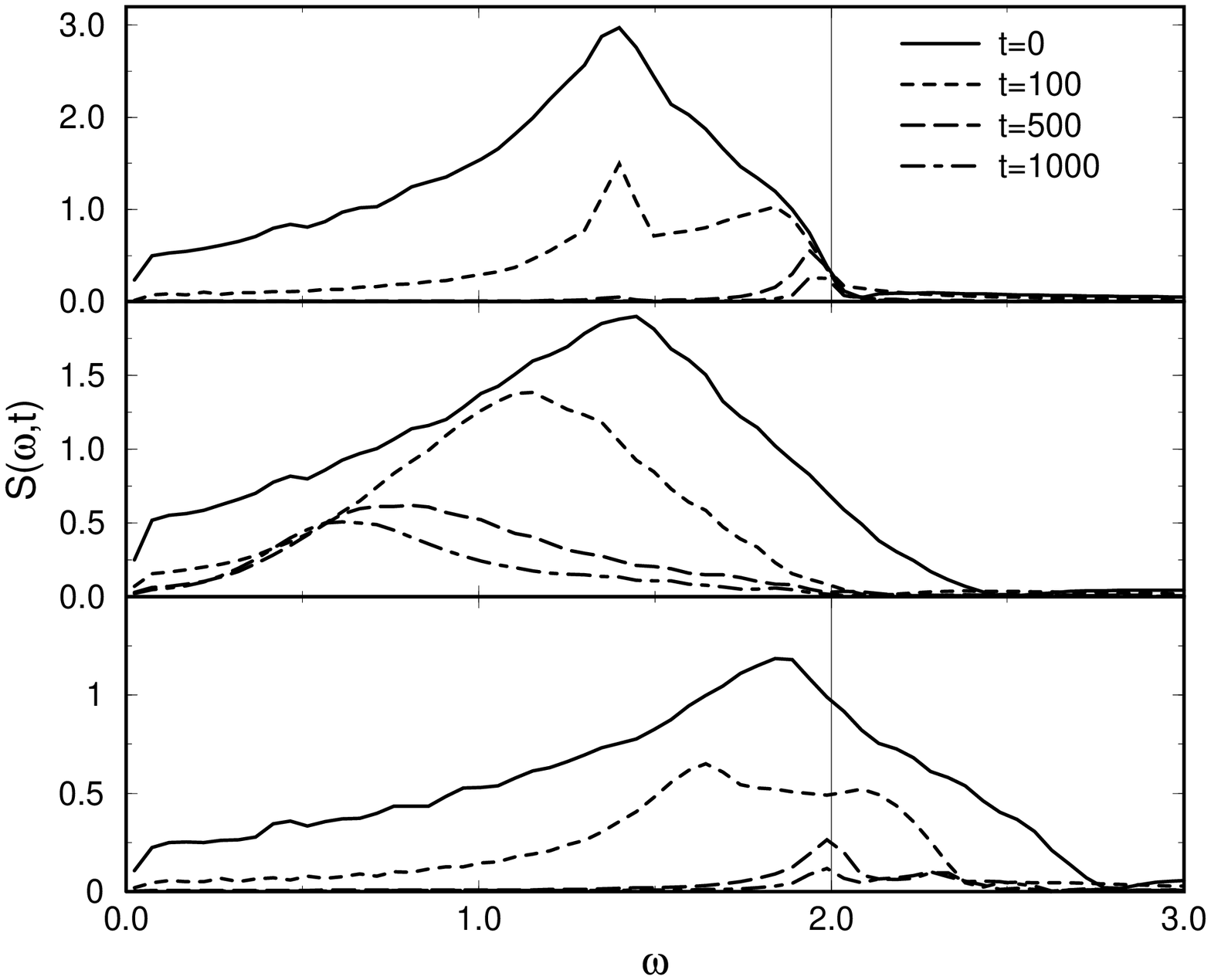}
\end{center}
\caption{Time evolution of spectra for various relaxing arrays. 
The upper three-frame panel is for one-dimensional ($50$ sites)
arrays, the lower three-frame panel for two-dimensional
($20\times 20$ sites) arrays. First
frames: harmonic interactions ($k=0.5$). Second frames: purely anharmonic
interactions ($k^\prime=0.5$).  Third frames: mixed interactions
($k=k^\prime=0.5$).  The time progression is as indicated.  The $t=0$
spectrum (solid curves) in each case is the equilibrium spectrum at
$T=0.5$, the initial temperature.  In all cases $\gamma=0.1$.  The thin
vertical lines indicate the frequencies $\omega=\sqrt{4k}=\sqrt{2}$
(1d panels) and $\omega=2$ (2d panels).
}
\label{spectrumdecay}
\end{figure}

The total energy of a relaxing array decays in time, and the questions of
interest are how exactly the energy decreases with time for arrays with
different interactions and in different dimensions.  The answers are
illustrated in Fig.~\ref{energydecay}.
Accompanying these decay curves are the more detailed spectral decay curves
shown in Fig.~\ref{spectrumdecay}.

The decay of the total energy ratio for harmonic systems is independent of
temperature, which is verified numerically and therefore leads to a
single curve in Fig.~\ref{energydecay} in each dimension for the given
parameters $k$, $N$, and $\gamma$.
The initial exponential decay of the energy in both one and two dimensions,
and the $N$ and $\gamma$ dependences of the decay rate, have been verified
numerically.  The long-time decay as an inverse power law
$E(t)/E(0) \sim t^{-\alpha}$ for both cases
has also been verified.
It is interesting to note that Eq.~(\ref{energy1}) can be rewritten
as an integral that makes explicit the cascade of relaxation times giving
rise to the inverse power law behavior:
\begin{equation}
\frac{E(t)}{E(0)} = e^{-t/\tau_0}I_0(t/\tau_0)=
\frac{\sqrt{\tau_0/2}}{\pi} \int_{\tau_0/2}^\infty d\tau ~
\frac{e^{-t/\tau}}{\tau\sqrt{\tau-\tau_0/2}}. 
\label{inverse}
\end{equation}
The lower limit, due to a nonzero shortest decay time, leads to the
initial exponential decay of the energy, but it is the long
$\tau^{-3/2}$ tail of slow relaxation times that leads to the power law
decay.  We return to this point below.

\begin{figure}[htb]
\begin{center}
\epsfxsize = 4.in
\epsffile{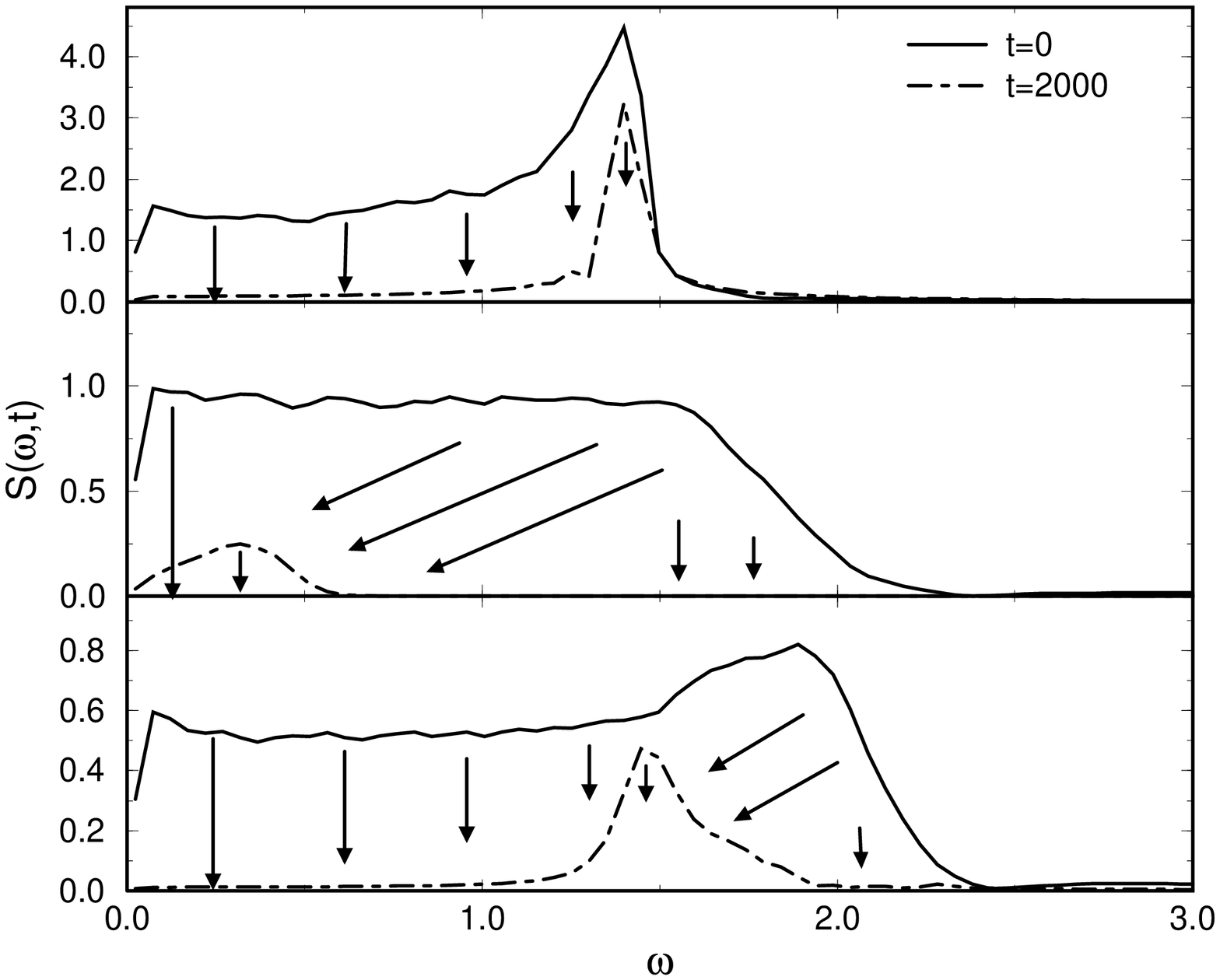}
\end{center}
\caption{Schematic representation of the spectral relaxation
channels in one dimension.  The one-dimensional spectra of
Fig.~\ref{spectrumdecay} for times $t=0$ and $t=2000$ are shown
again here, and the
arrows depict the pathways of different spectral components.  Downward
arrows indicate absorption by the cold reservoir, while angled arrows
denote degradation from one spectral region to another.  The
relative lengths of the arrows depict the associated rates.}
\label{figswesq}
\end{figure}

The associated spectra for the harmonic arrays are shown in each of the
first frames of the two panels in Fig.~\ref{spectrumdecay}.   
The evolution confirms that {\em low} frequencies decay more
rapidly in the harmonic chain -- the spectrum is
absorbed by the cold reservoir from the bottom up, and by the
latest times shown, only the longer-lived band-edge modes remain
in the system.  Since each spectral component is associated with an
independent phonon, the
spectral decrease occurs ``vertically", that is, each spectral
component decays directly into the reservoir on its characteristic time
scale; this is shown schematically for the one-dimensional systems
in Fig.~\ref{figswesq}, where the downward arrows represent absorption by
the reservoir and their relative length schematizes the absorption rate.

For anharmonic arrays the relaxation behavior depends strongly on the
presence or absence of a harmonic component in the potential, and, in some
respects, on dimensionality.  First we consider the purely anharmonic arrays.
The dimensionality plays a particularly important role in this case.

In one dimension the upper panel of Fig.~\ref{energydecay} shows an
essentially purely exponential decay (verified separately),
which is characteristic of a single predominant decay channel.  The decay
is more rapid at higher temperatures.  Note that there are no phonons in
this purely quartic system, so that single frequencies are not associated
with normal modes of the system.  Conversely, exact solutions of the
FPU chain such as solitons and intrinsic localized modes may involve
many frequencies.
The second frame in the upper panel of Fig.~\ref{spectrumdecay} shows that
the higher frequencies decay first, exactly opposite to the harmonic chain.
We find that the relaxation pathway is for the high frequency portions of
the spectrum to degrade rapidly into lower frequency excitations,
as schematically indicated by the sloped arrows in Fig.~\ref{figswesq}. 
The lowest frequencies decay into the reservoir and define the
exponential decay rate seen in Fig.~\ref{energydecay}. 
In more detail, the high frequency components of the spectrum
are mainly due to highly mobile
localized modes that degrade into lower energy (less mobile) excitations
as they move and collide with one another.
The lowest frequency excitations are in turn absorbed
into the cold reservoir but continue to be replenished through the
degradation process.   It is important to note, however, that
among the low frequency excitations are some
that persist for a very long time, certainly beyond the times of
our simulations.  Their decay is surely slower than exponential, perhaps a
stretched exponential.  These, the only remaining spectral components
at time $t=2000$,
are ``labeled" by short downward arrows in the relaxation
schematic and include rather stable breather
and/or soliton modes that move very slowly and are localized away from the
boundaries.  Also, some direct
relaxation of all frequency components into the reservoir occurs as well
(shown schematically by the short arrows at high frequencies in
Fig.~\ref{figswesq}), but this direct relaxation is slower than the energy
degradation pathway. For example,
when a highly mobile localized excitation reaches a boundary, it
typically remains at the boundary for about one period of oscillation
(which is short for a highly energetic excitation),
during which it loses a small portion of its energy to the reservoir.
The remaining excitation is reflected back into
the chain, where it will continue to lose energy through other collision
events and/or re-arrival at the boundaries.
The role of high-frequency mobile
modes and of low-frequency slowly moving or stationary modes
in this picture will be tested in more detail in the next section,
where we explicitly inject a high-frequency localized mode into the array
and observe the relaxation
dynamics. We do note here that our picture is consistent with
known facts about localized states.  In particular, it is known that
higher-frequency and/or higher amplitude localized modes can
move at higher velocities~\cite{bourbonnais,cretegny,kosevich}.
It is also known that while in motion such modes lose energy through
collisions with other excitations.
Figure~\ref{energydecay} shows a faster decay at higher temperatures,
which is consistent with our observations elsewhere that the speed
of an injected pulse (and therefore, we conjecture, the speed of a
moving localized mode) in these arrays increases with
temperature~\cite{wepulse}.   

\begin{figure}[htb]
\begin{center}
\leavevmode
\epsfxsize = 4.in
\epsffile{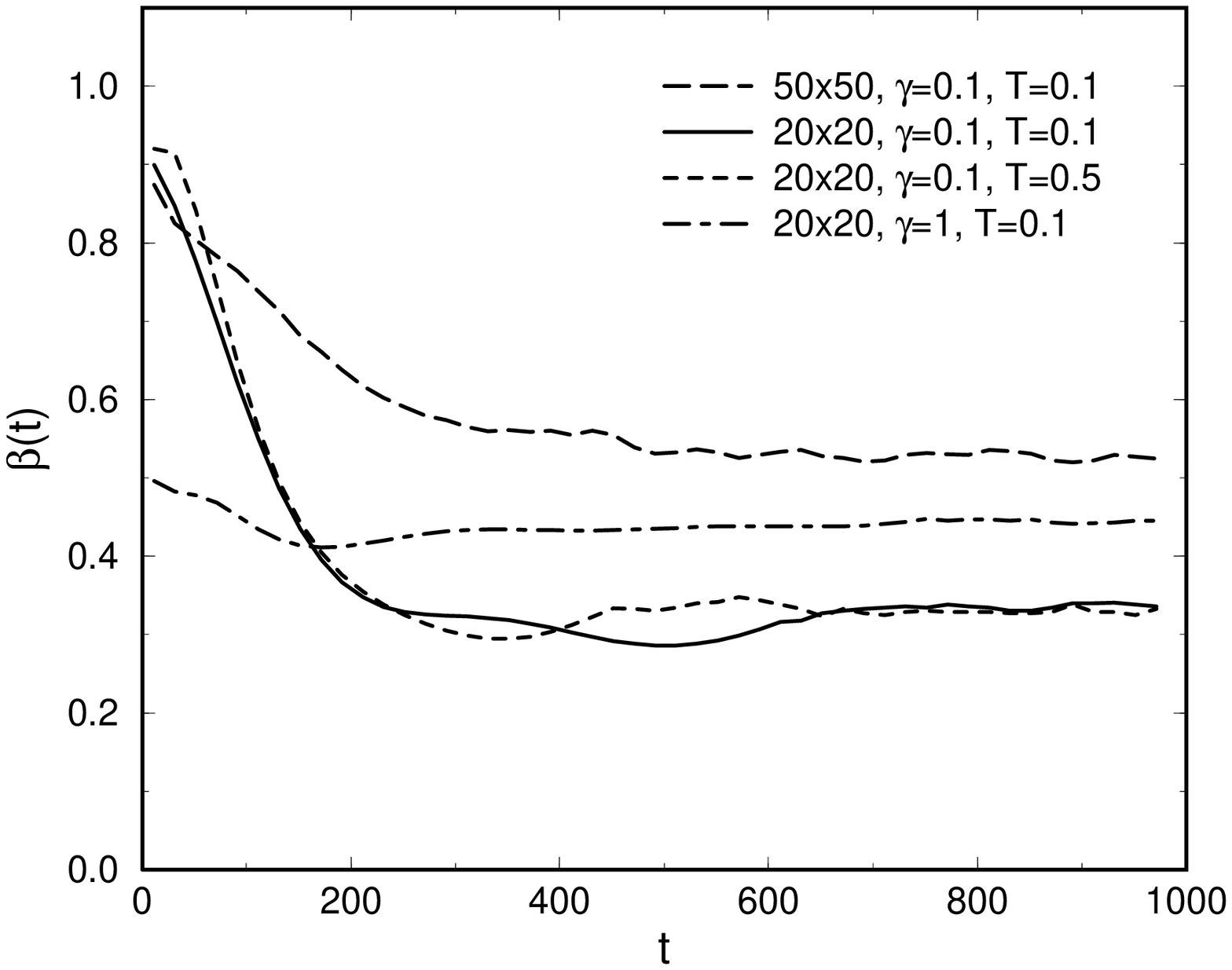}
\leavevmode
\epsfxsize = 4.in
\epsffile{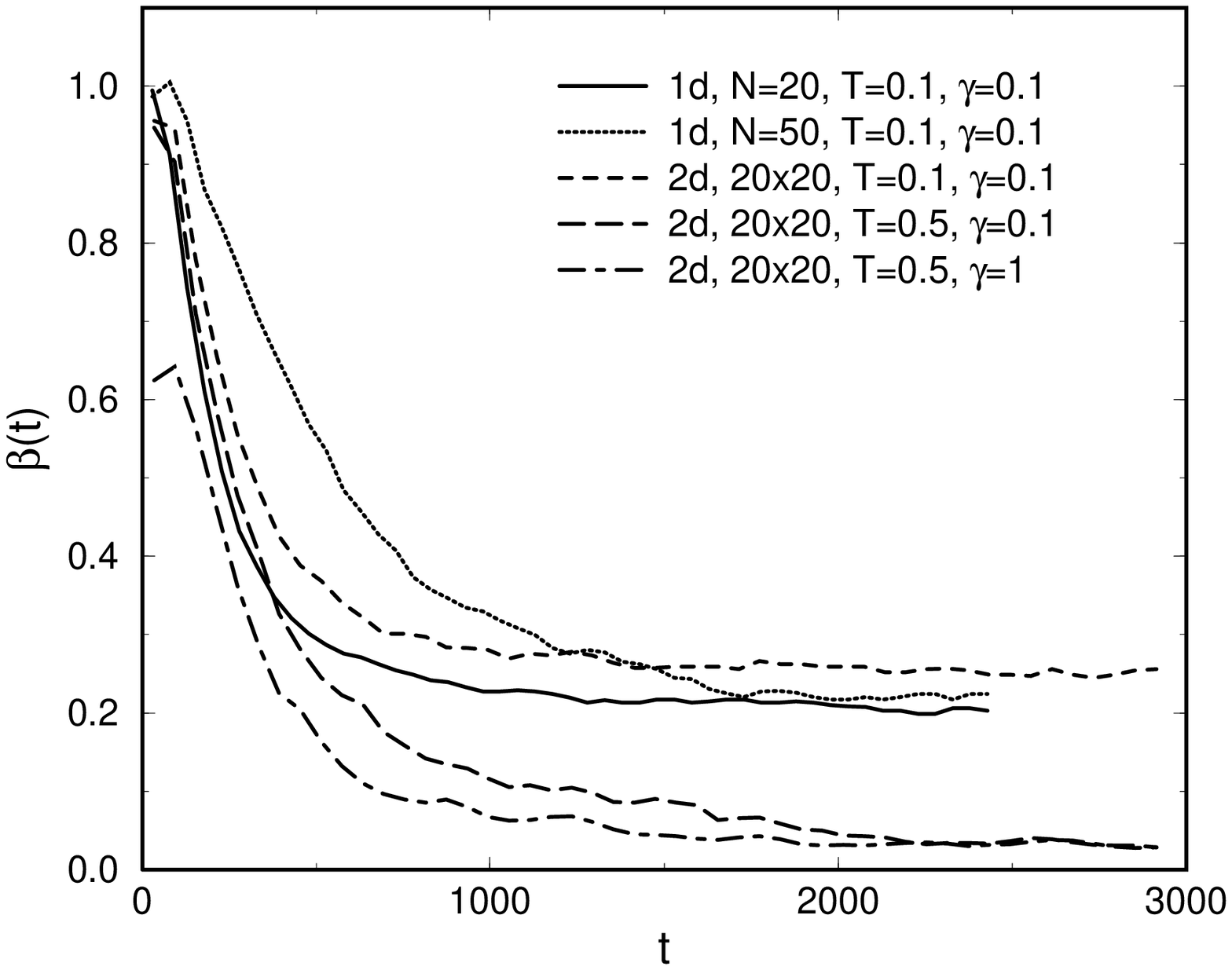}
\end{center}
\caption{Plot of $\beta(t)$ as a function of time.  A flat line
below $\beta(t)=1$ indicates stretched exponential behavior.
First panel: purely anharmonic 2d arrays of different sizes, damping
coefficients, and temperatures.  Second panel: various 1d and 2d mixed
arrays.
}
\label{figexp}
\end{figure}

The relaxation dynamics of the purely anharmonic array in two dimensions 
differs from the one-dimensional case in a number of ways.
First, we note that the decay of the energy in Fig.~\ref{energydecay}
is (except for very early times) slower in the hard array than in
the harmonic one over the times of observation.  Second, except
for an initial short time interval, the decay is found not
to be exponential, indicating that there is no longer a single
predominant decay channel as there was in one dimension.  In two dimensions
the relaxation pathway again includes degradation of higher frequency
excitations to lower frequencies, as can be seen in the spectral rendition
in Fig.~\ref{spectrumdecay}.  However, whereas in one dimension the
degradation process is faster than the decay of low frequency
excitations into the reservoir (and hence this
latter decay is the rate-limiting step that defines the
total energy decay process), here the degradation process is slower, leading
to spectral bottlenecks and competing time scales.
We find that increasing the array size leads to slower degradation of the
high frequency components and to more pronounced spectral bottlenecks
in the mid-frequency range.  Our physical picture of the source of the
competing time scales involves the observation that in two
dimensions, localized excitations are not nearly as mobile as in one
dimension.  In one dimension energy
degradation occurs as a consequence of high mobility and the
resultant inevitable frequent scattering.  The reduced mobility
in two dimensions was noted
in earlier work on pulse propagation~\cite{wepulse} and will be supported
and clarified in the next section, where we explicitly inject a 
high-amplitude breather into our system.  
At long times in the 2d system excitations of all energies
eventually reach the
boundaries.  These excitations typically lose some of their energy into the
cold reservoir and the remainder is reflected back into the array, where it
either degrades into lower energy components or reaches a boundary
again with the attendant energy loss.  Note also that with increasing
temperature the total system energy decays more rapidly, which is
consistent with our assertion that mobility (low as it may be) in the
purely hard arrays increases with temperature because of the participation
of more mobile higher frequency components in the initial equilibrium
mix of excitations.

The result is an energy decay that is of stretched exponential form,
\begin{equation}
\frac{E(t)}{E(0)} \sim e^{-(t/\tau_1)^\sigma},
\label{stretched0}
\end{equation}
as is evident in the first panel of Fig.~\ref{figexp}, where we plot
\begin{equation}
\beta(t) = \frac{d}{d\ln t} \ln \left[ -\ln
\left(\frac{E(t)}{E(0)}\right)\right]
\end{equation}
as a function of time for four
purely anharmonic 2d arrays of different sizes, damping coefficients, and
temperatures.  When the decay is a stretched
exponential this rendition gives a flat line at the value
$\beta(t)=\sigma$.  
We find that the value of $\sigma$ is independent of temperature,
as seen in the figure
(we have tested this assertion for various lattice sizes in the
range $T=0.1-1.0$), and that it increases with
$\gamma$, as also seen in the figure.  For the
$20\times 20$ lattices we find that $\sigma$ is around
$0.33$ for $\gamma=0.1$ and $\sigma\approx 0.43$ for $\gamma=1$. 
The temperature independence is explained by the fact that 
temperature does not dominate the mobility of the
residual excitations nor does it determine their rate of energy loss
once they reach a boundary.  On the other hand, it is reasonable that
increasing $\gamma$ leads to a faster long-time relaxation process (larger
$\sigma$) because more energy is lost to the reservoir upon each collision.
We find that $\sigma$ also increases with $N$, which is somewhat of a
puzzle.  We find, for instance,
that in a $50 \times 50$ array with $\gamma=0.1$ (and for any temperature)
$\sigma \approx 0.52$.  The issue of the size dependence of relaxational
processes following a stretched exponential behavior is a difficult problem
that has only recently been addressed in a different context~\cite{bunde}.
In summary, for the purely anharmonic 2d arrays
\begin{equation}
\sigma = \sigma(N,\gamma).
\end{equation}

The mixed arrays, i.e., those with interactions that have both quadratic and
quartic potential contributions, are of course the ubiquitous FPU systems 
since it is difficult to envision a ``real" physical system that has no
quadratic potential terms (one might also say this about cubic potential
terms that have not been included here~\cite{future}). 
The thermal relaxation of these arrays proceeds
similarly in one and two dimensions.  At early and intermediate times
the mixed arrays relax very similarly to the harmonic arrays (i.e.,
exponential decay followed by power law decay), albeit
somewhat modified and speeded up by the presence of
high-frequency mobile excitations in addition to the low-frequency
phononic excitations.  The rapid decay of both low and high
frequency excitations is evident in the third frames of both panels in
Fig.~\ref{spectrumdecay}.  This similarity in one dimension was 
noted by Piazza et al.~\cite{piazza}.  
After some time, however, the mixed chain relaxation behavior changes
to a stretched exponential.  This occurs when the low frequency
modes have essentially all decayed.  The higher frequency spectral
components that persist are localized long-lived excitations
(note that high-frequency phonon modes are unstable against breather
formation in these systems~\cite{cretegny}).  Unlike the purely anharmonic
array, here there is no low-frequency residue to perturb the high-frequency
localized excitations and so they survive relatively unperturbed
and immobile for a long time. 
The persistence of high-frequency spectral components
is seen in the third frames of both panels in Fig.~\ref{spectrumdecay}, and
the schematic representation of the progression in one dimension is shown
in Fig.~\ref{figswesq}.  The slow leakage of breather energy into
low energy modes that continue to dissipate into the cold reservoir is
responsible for the eventual stretched exponential relaxation of the
system.

These behaviors can be seen clearly in the second panel of
Fig.~\ref{figexp} which shows
$\beta(t)$ for various 1d and 2d mixed arrays.
The initial exponential behavior (which corresponds to a flat curve at
$\beta(t)=1$), occurs over too short a time scale to be clearly
discernible.  There then follows a
power law decay regime which eventually turns to a stretched exponential.
In the power law regime, if the energy decays as $E(t)/E(0) \sim
(t/\tau_0)^{-\alpha}$ then it follows that $\beta(t)= \left(\ln
\frac{\tau}{\tau_0}\right)^{-1}$, which is independent of $\alpha$ and
depends only on the ratio $N/\gamma$ via $\tau_0$.  The two 2d curves
and the 1d curve with the same value of $N/\gamma$
are all seen to decay with the same slope.  As all the
curves settle into their asymptotic behavior, we see that $\sigma$ 
depends neither on $N$ nor on $\gamma$: the two 1d arrays are
of different sizes but asymptote to the same $\sigma$ (approximately
$0.21$), as do the two 2d
arrays of the same size and initial temperature but with different 
damping coefficients ($\sigma \approx 0.03$).  On the other hand,
the 2d arrays that have different initial temperatures asymptote to
different values of $\sigma$ (approximately $0.25$ for $T=0.1$ and
around $0.03$ for $T=0.5$). 
The observed behavior is
explained by the fact that the slow leakage of energy out of the long-lived
localized excitations is rate limiting.  A higher initial temperature
leads to more energetic, more stable breathers with slower leakage, hence
explaining why $\sigma$ decreases with increasing temperature.  Neither
the size of the system nor the damping coefficient are important in this
limit, since the slowest process is the leakage.  The low-energy
outcome of that leakage is absorbed quickly by the cold reservoir.
Note that this description is consistent with that provided for
relaxation of lattices with local anharmonic potentials~\cite{tsi2}.
In summary, for 1d and 2d arrays with quadratic plus quartic
interaction potentials
\begin{equation}
\sigma = \sigma(T).
\end{equation}

It is interesting to stress that, like an inverse power law decay,
a stretched exponential can indeed be obtained from a distribution
or hierarchical progression of decay times.  For example, 
\begin{equation}
e^{-(t/\tau_1)^{1/2}} 
=\frac{1}{2\sqrt{\pi\tau_1}}
\int_0^\infty d\tau ~ e^{-t/\tau} ~ \frac{e^{-\tau/4\tau_1}}{\tau^{1/2}},
\label{stretched}
\end{equation}
which should be compared with Eq.~(\ref{inverse}). 
More generally, if the distribution of relaxation times varies as
$e^{-(\tau/\tau_1)^\mu}$, then the decay will go as a stretched
exponential with $\sigma\sim 1/(1+\mu)$ at long times
(the stretched
exponential with $\sigma=1/2$ is the only one whose associated distribution
is expressible in extremely simple analytic form, but distributions
associated with other fractional exponents are also known analytically
or numerically~\cite{montroll}). 
The distributions leading to inverse power decay and to a stretched
exponential decay are both broad, but the inverse power law
of course arises explicitly from very
long tails not present in the stretched exponential.
In other words,
the relaxation of the last energy residues of a very large
harmonic lattice take {\em longer}
than those of a very large anharmonic lattice.  There is an important
difference, however: in the harmonic lattice the persistent excitations are
distributed over the entire lattice, while in the anharmonic lattice they
are localized.  Also, in a {\em finite} lattice eventually the harmonic
decay is again exponential while that of the anharmonic system remains
a stretched exponential. 

\begin{figure}[htb]
\begin{center}
\epsfxsize = 1.7in
\epsffile{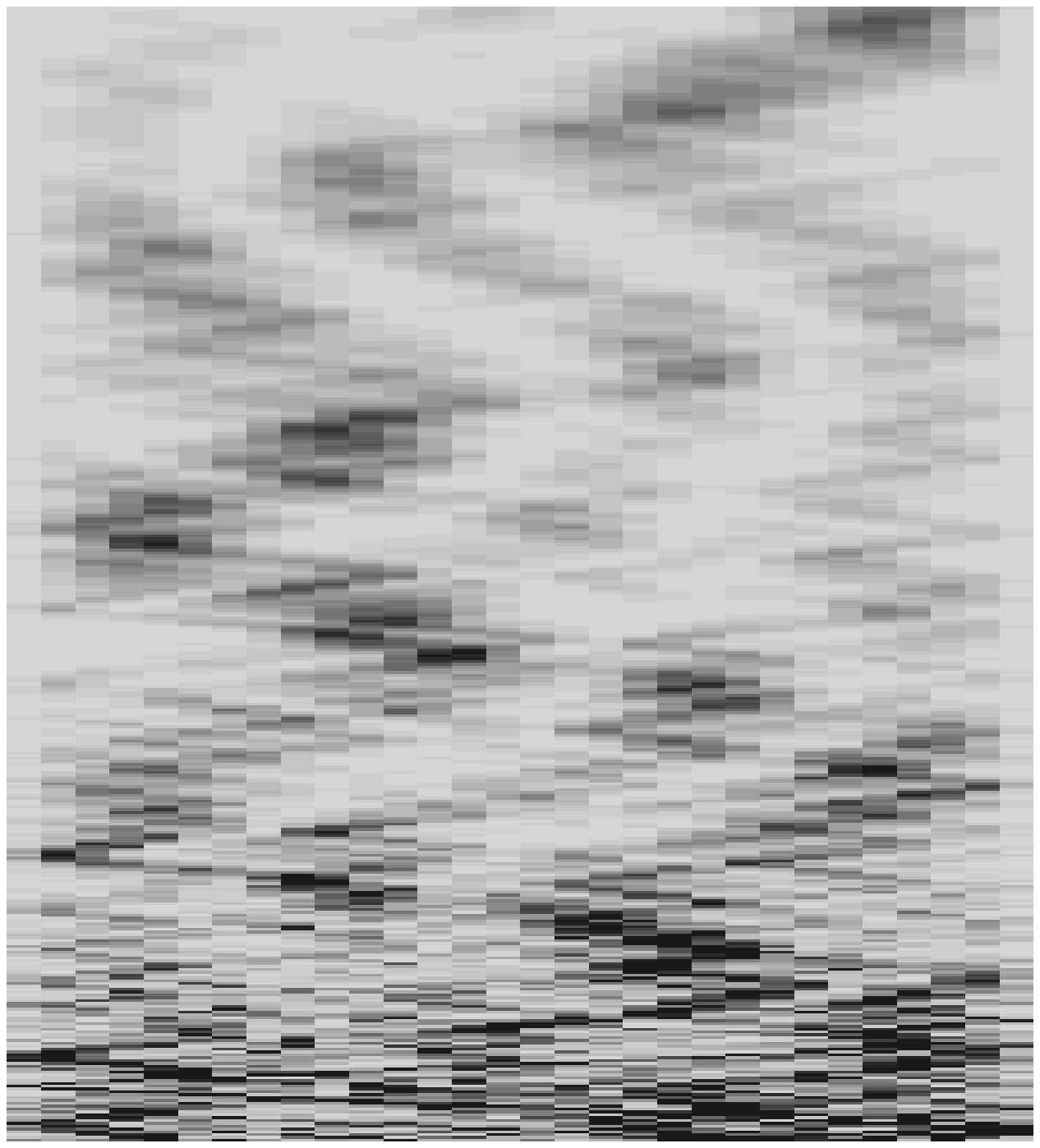}
\epsfxsize = 1.7in
\epsffile{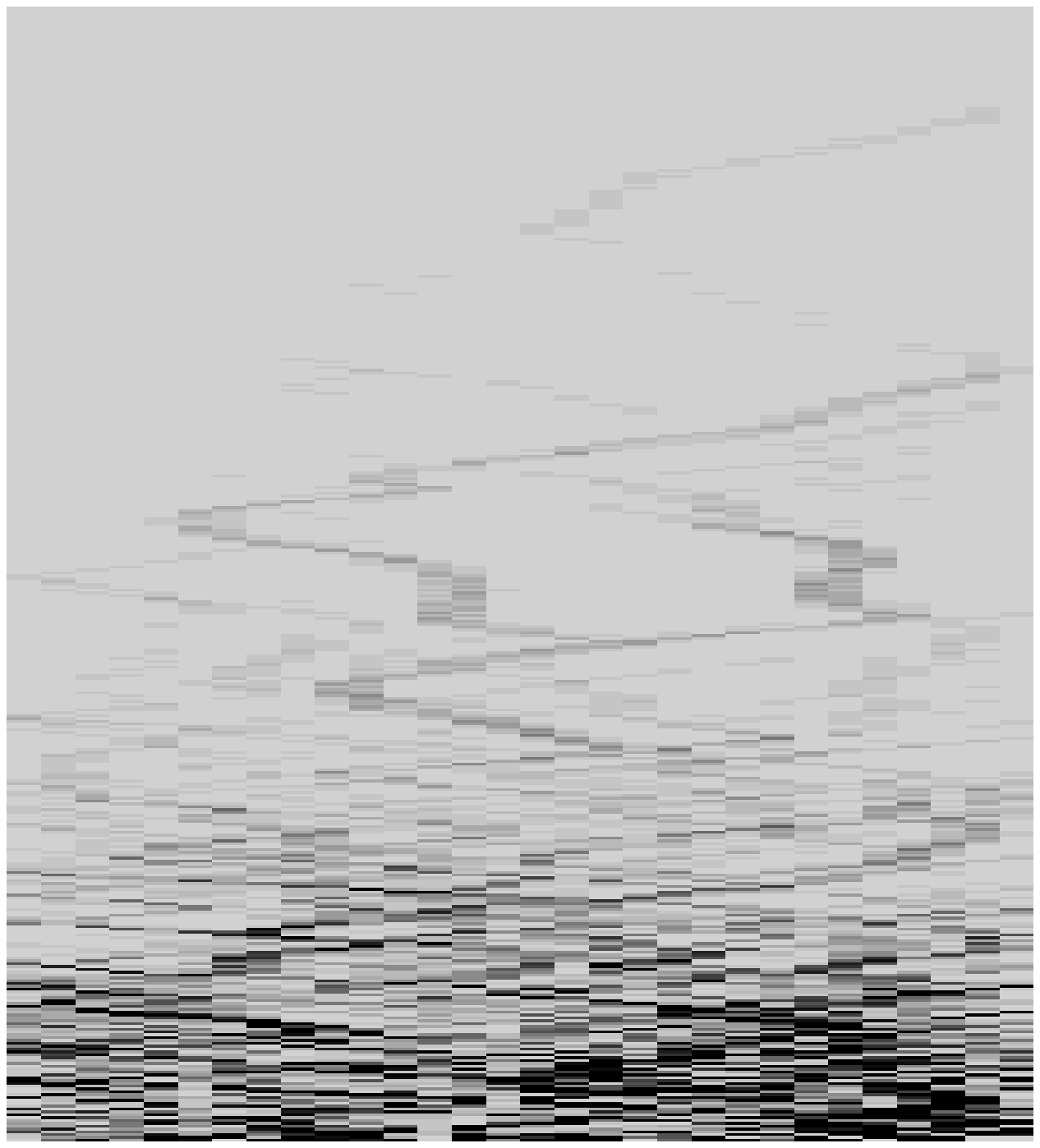}
\epsfxsize = 1.7in
\epsffile{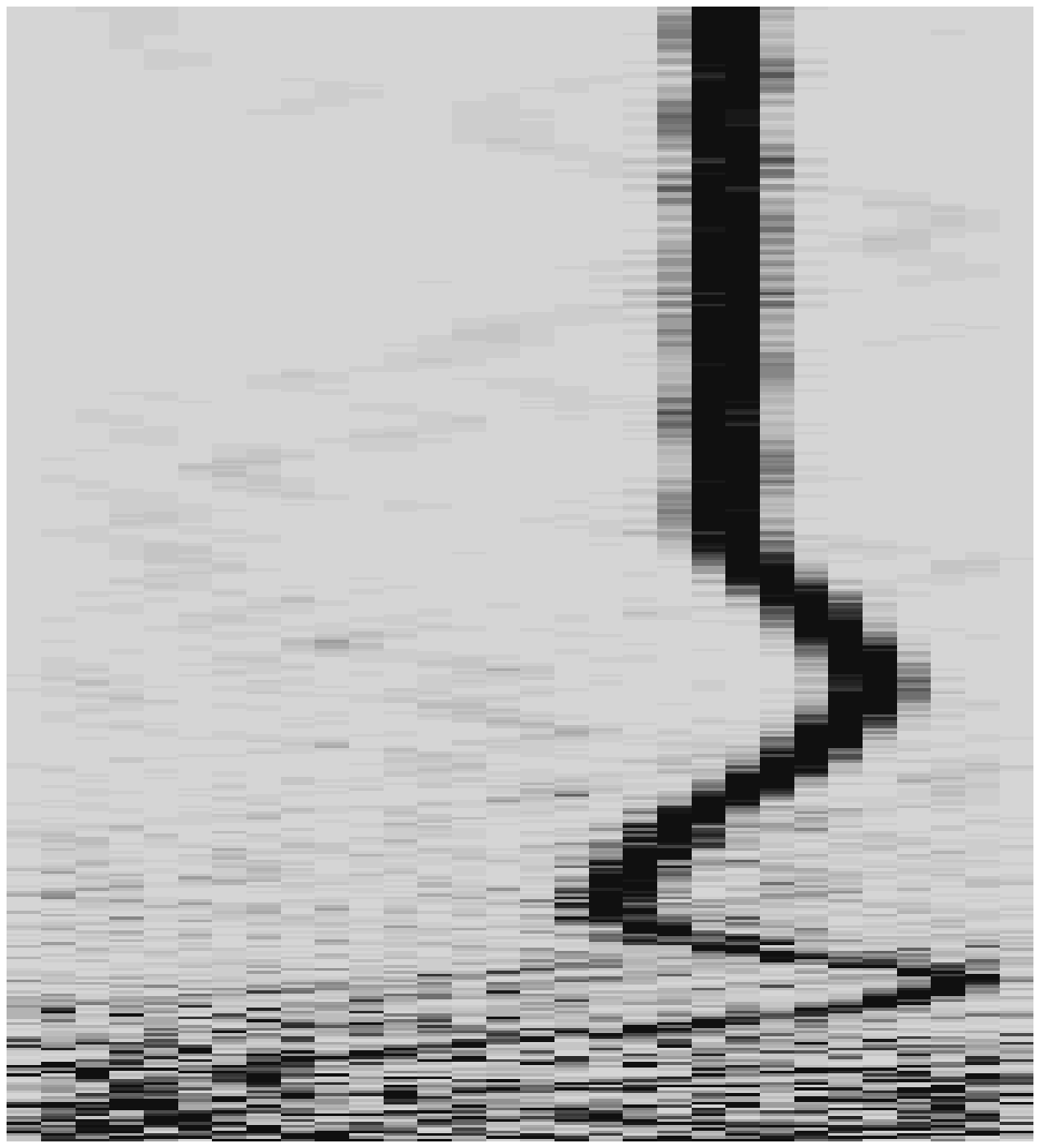}
\end{center}
\caption
{Local energy landscapes of one-dimensional 30-site arrays initially
thermalized at $T=0.5$.
Time advances along the $y$-axis until $t=1000$.
A gray scale is used to represent the local energy,
with darker shading corresponding to more energetic regions. First panel:
harmonic
chain, $k=0.5$ and $k^\prime=0$.  Second panel: purely anharmonic chain, $k=0$
and $k^\prime=0.5$.
Third panel: mixed chain, $k=k^\prime=0.5$.}
\label{figevol}
\end{figure}

\begin{figure}[htb]
\begin{center}
\includegraphics[height=0.15\textheight]{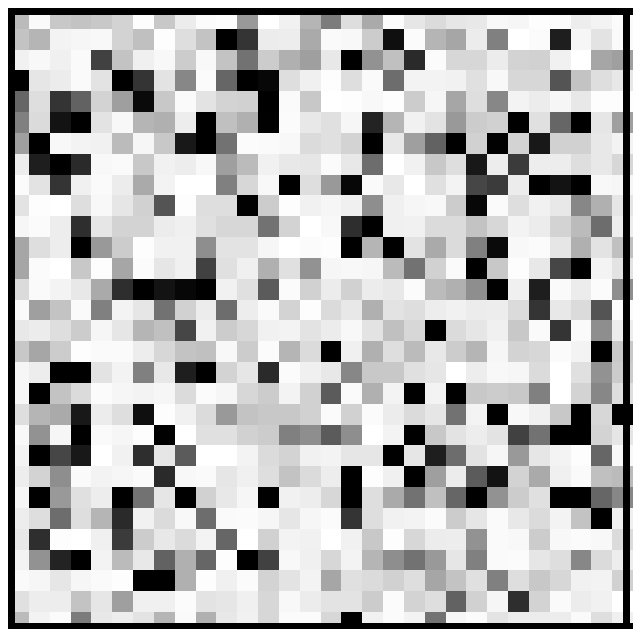}
\includegraphics[height=0.15\textheight]{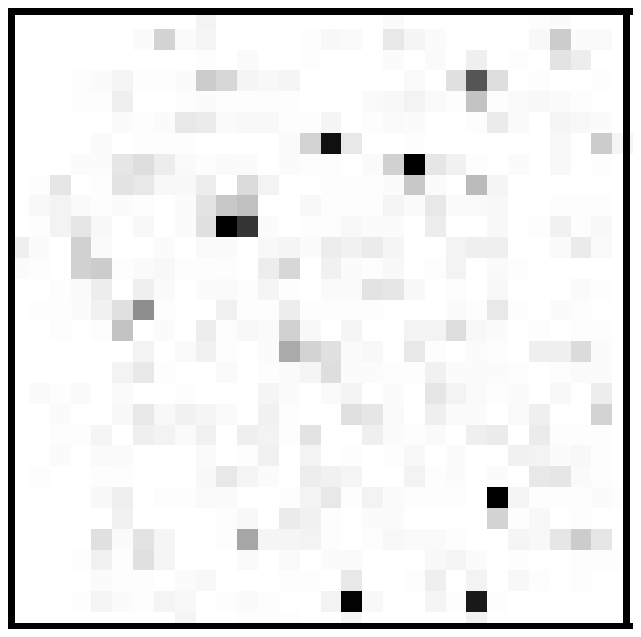}
\includegraphics[height=0.15\textheight]{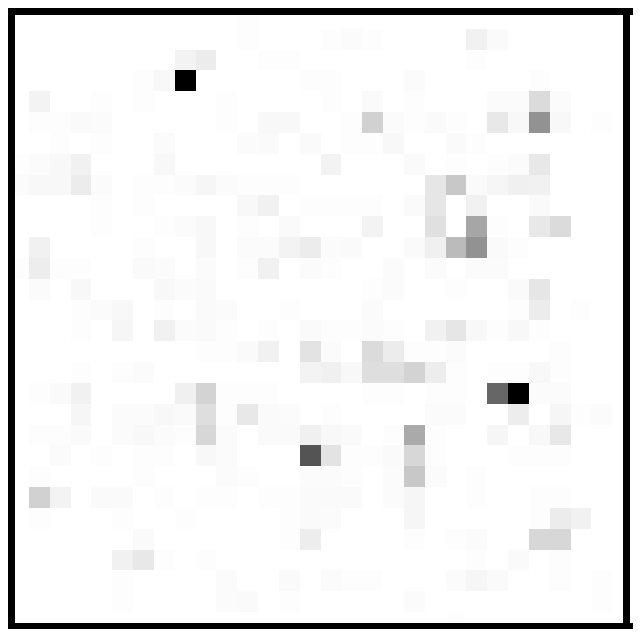}
\includegraphics[height=0.15\textheight]{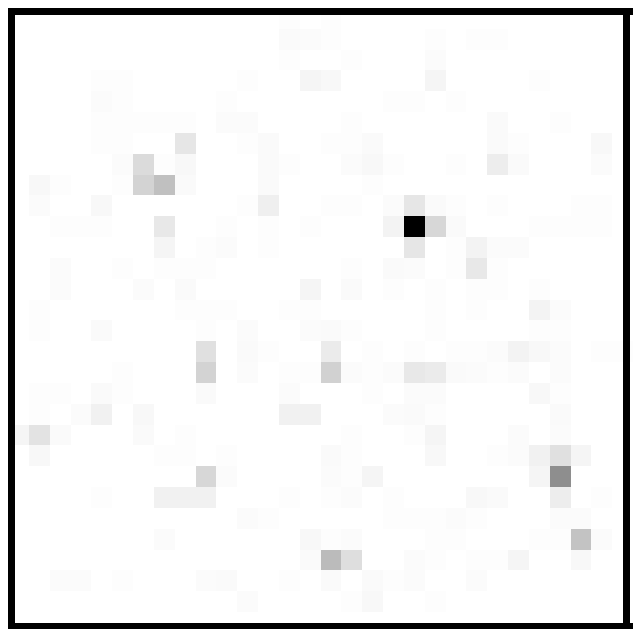}
\includegraphics[height=0.15\textheight]{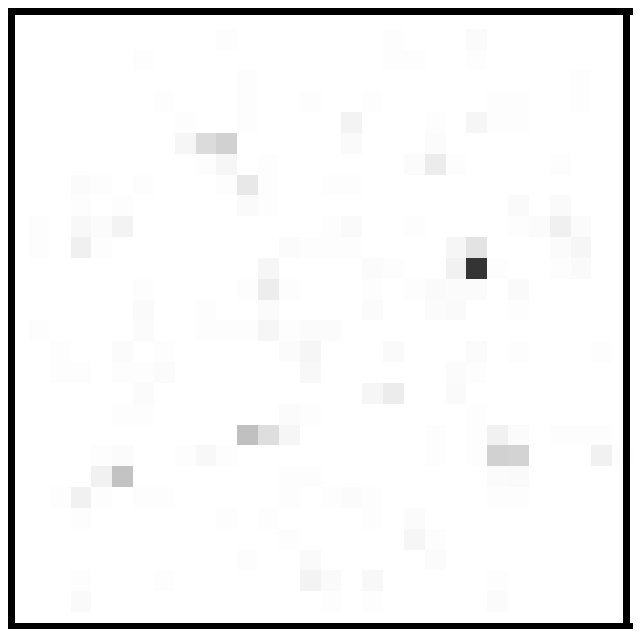}
\includegraphics[height=0.15\textheight]{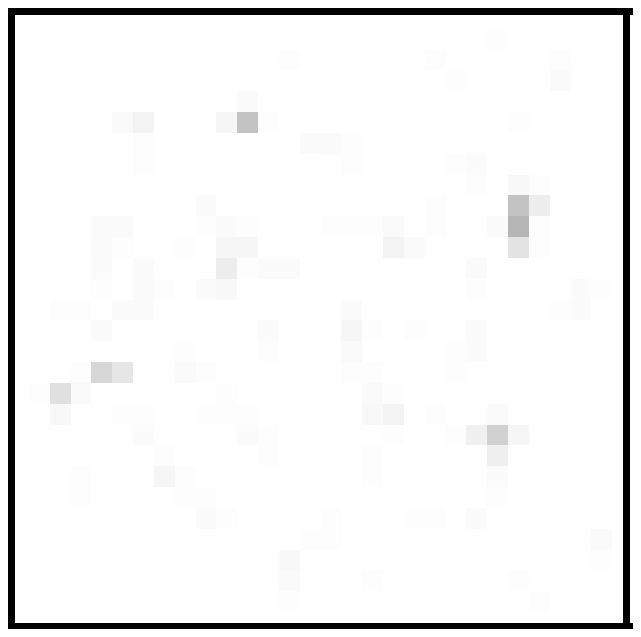}
\end{center}
\caption
{Local energy landscapes for a $20\times 20$ purely anharmonic lattice
($k^\prime=0.5$) initially thermalized at $T=0.5$ and with $\gamma=0.1$.
From first to last frames: $t=0$, $400$, $800$, $1200$,
$1600$, and $2000$.}
\label{figevold2}
\end{figure}

\begin{figure}[htb]
\begin{center}
\includegraphics[height=0.15\textheight]{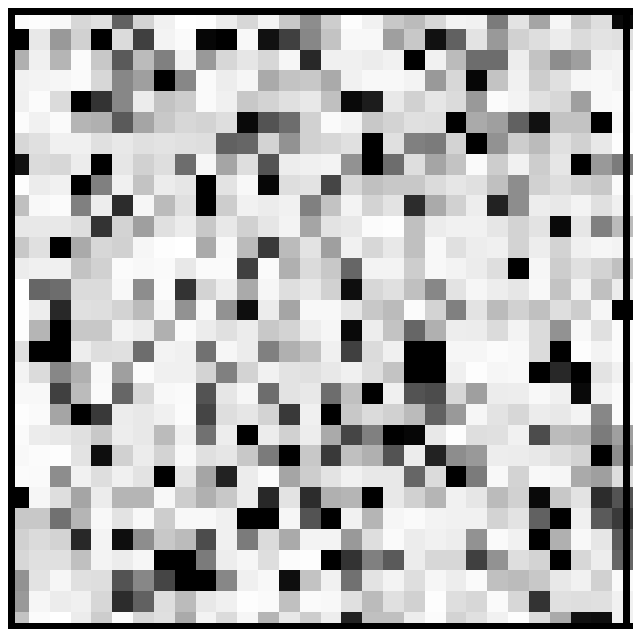}
\includegraphics[height=0.15\textheight]{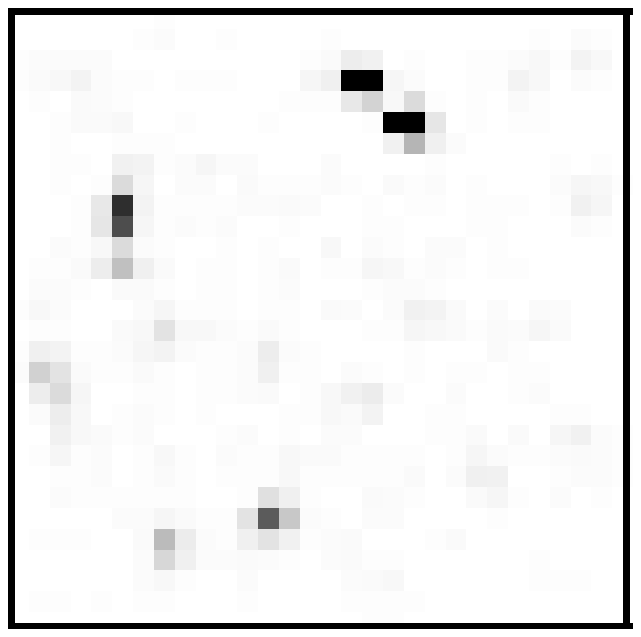}
\includegraphics[height=0.15\textheight]{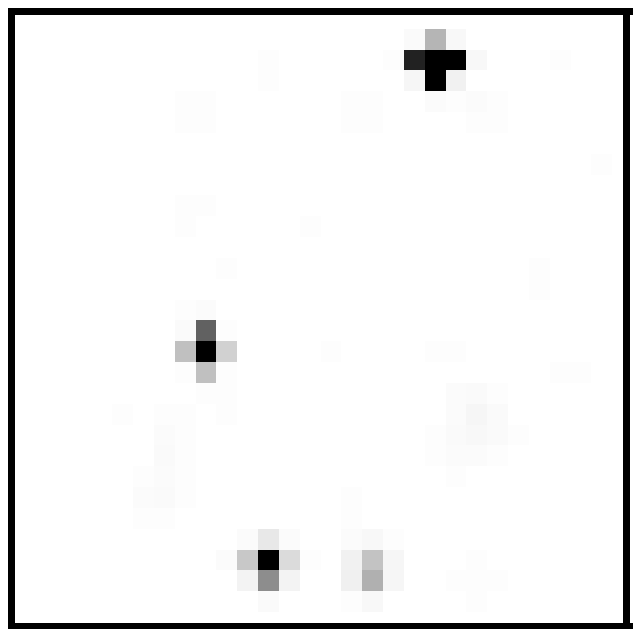}
\includegraphics[height=0.15\textheight]{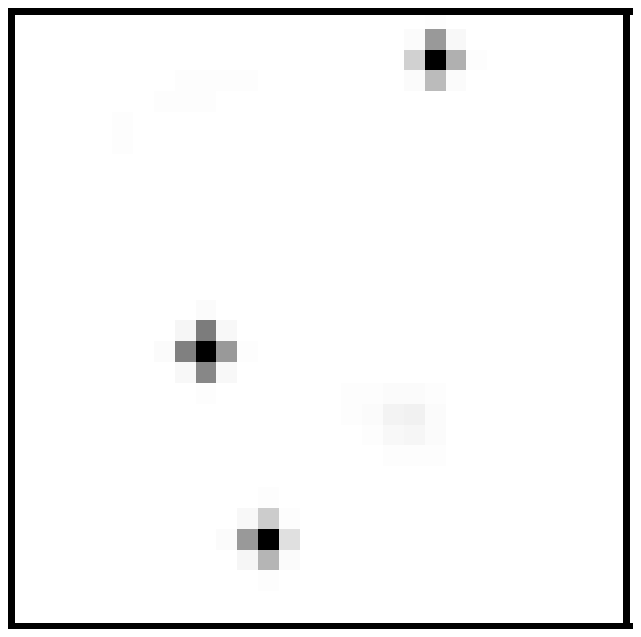}
\includegraphics[height=0.15\textheight]{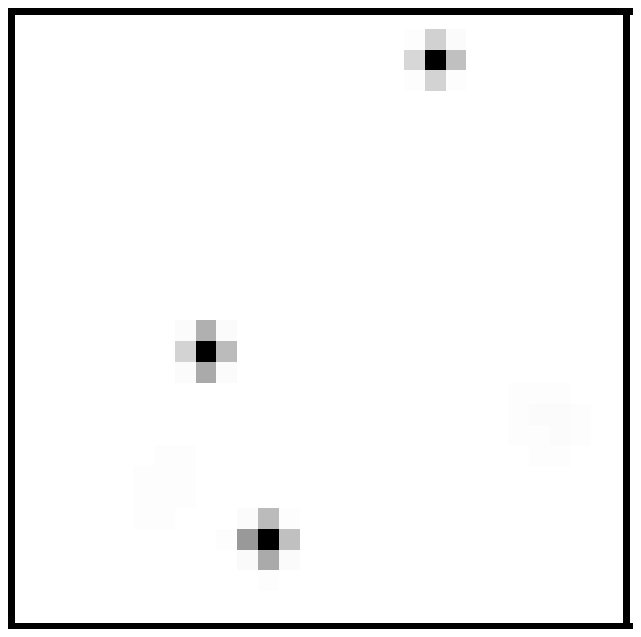}
\includegraphics[height=0.15\textheight]{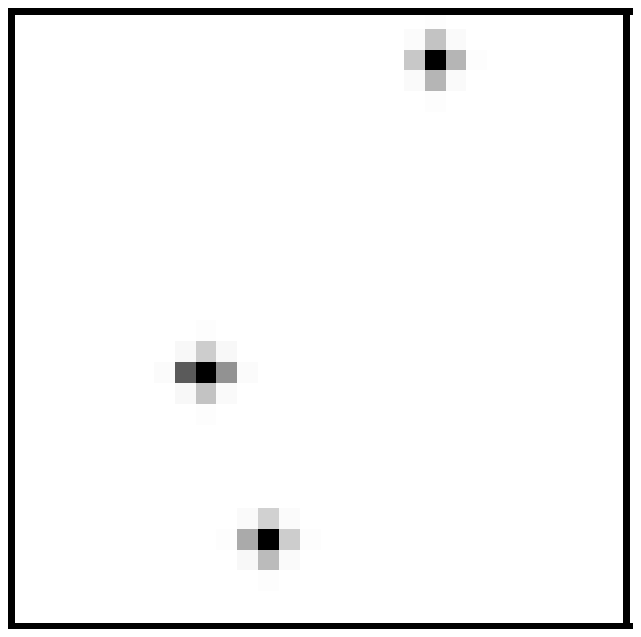}
\end{center}
\caption
{Local energy landscapes for a $20\times 20$ mixed anharmonic lattice
($k=k^\prime=0.5$) initially thermalized at $T=0.5$ and with $\gamma=0.1$.
From first to last frames: $t=0$, $400$, $800$, $1200$,
$1600$, and $2000$.}
\label{figevolhhdd}
\end{figure}

Spatial energy landscapes rendered in gray scale provide a pictorial 
representation of the thermalization progressions.  
We follow widespread convention and define the local energy in one
dimension as
\begin{equation}
E_i=\frac{\dot{x}_i^2}{2} +\frac{1}{2}\left[ V(x_i-x_{i-1}) +
V(x_{i+1}-x_i)\right]
\label{local}
\end{equation}
with obvious generalization in two dimensions. 
Figure~\ref{figevol} shows the temporal evolution of the local
energy landscape 
for each of the three chains.  Particularly dramatic is the spontaneous
occurrence of an essentially stationary breather in the mixed chain.  It is
this sort of breather that leads to the extremely slow relaxation of the
mixed chain energy.  Spatial landscapes for two-dimensional systems are
shown in six time frames in Figs.~\ref{figevold2} and \ref{figevolhhdd}
for the pure anharmonic and the mixed anharmonic lattices, respectively.
Note that in the purely hard array the localized high-energy regions move
around and relax within the time scale of the progression.  In the mixed
array, on the other hand, the ``hot spots" persist and essentially do
not move. 

\section{Relaxation With an Injected Breather}
\label{breather}

Our portrayal of the thermal relaxation process can be further bolstered by
initially injecting a high-energy localized excitation in the center of
each thermalized array and observing the behavior of this excitation during
the relaxation process.  Some of the dimensionality differences are thereby
clarified.  The injected excitation is chosen so as to be close to a
known exact breather solution of the anharmonic arrays.

In a one-dimensional array with an interaction potential $V(x_i-x_{i-1}) =
(x_i-x_{i-1})^n$ as $n\longrightarrow\infty$ and exact odd-parity
breather is one of amplitude $A$ at a site and $-A/2$ at each
of the two immediately adjacent sites. 
An exact even-parity breather is one with amplitude $A$ at
one site and $-A$ at an immediately adjacent site.  These are not exact
solutions when $n$ is not infinite and/or when there are quadratic
contributions to the potential, but they are close to exact, even for
the FPU chain~\cite{flach,MA}.  In two dimensions the odd-parity solution
with amplitude $A$ at one site and amplitudes $-A/4$ at the four nearest
neighboring sites is also nearly exact, but there is no equivalent to the
even-parity breather.  We insert an odd-parity excitation at the center of
our array and in each case choose arrays sufficiently large ($300$ sites in
one dimension, $30\times 30$ in two dimensions) so that the excitation
either decays or stops moving before ever reaching the boundaries.  In
harmonic arrays the fate of such an injected excitation is completely
predictable and uninteresting: it spreads quickly over the entire array and
thus loses its localized character.  The associated Fourier decomposition
into phonon modes dictates the relaxation behavior. 

To follow the excitation in the one-dimensional anharmonic arrays
we calculate the mean squared displacement
\begin{equation}
\left< x^2(t)\right> \equiv \left< (i_{max}(t) - \frac{N}{2})^2\right>
\end{equation}
as a measure of the position of the excitation
(its dispersion in the anharmonic chains is very
small~\cite{wepulse}).  Here $N/2$ is the initial point of
highest energy in the chain and $i_{max}(t)$ is the point of maximum energy
at time $t$.  Similarly, in two dimensions we define
\begin{equation}
\left< r^2(t)\right> \equiv \left< (i_{x,max}(t) - \frac{N}{2})^2\right> 
+ \left< (i_{y,max}(t) - \frac{N}{2})^2\right>
\end{equation}
where $(N/2,N/2)$ is the initial point of highest energy and
$\left(i_{x,max}(t),i_{y,max}(t)\right)$ is the point of
maximum energy at time $t$.  

In a purely anharmonic chain in one dimension, a short time after it is
created the excitation begins to move
essentially ballistically in one direction or the other with
equal probability.  The motion continues for a period of random duration,
until the excitation stops moving for a random period of time.  Then
it moves again in either direction.  Whatever its initial parity,
while subsequently stationary the excitation has even parity (the
more stable of the two configurations).  Any perturbation (usually
scattering of slow low-frequency excitations) that disturbs this
parity sets the breather in motion, and while it moves it alternates
between even and odd parity. The excitation only loses energy
while in motion, through collisions with persistent low-frequency
excitations. 

\begin{figure}[htb]
\begin{center}
\epsfxsize = 4.in
\epsffile{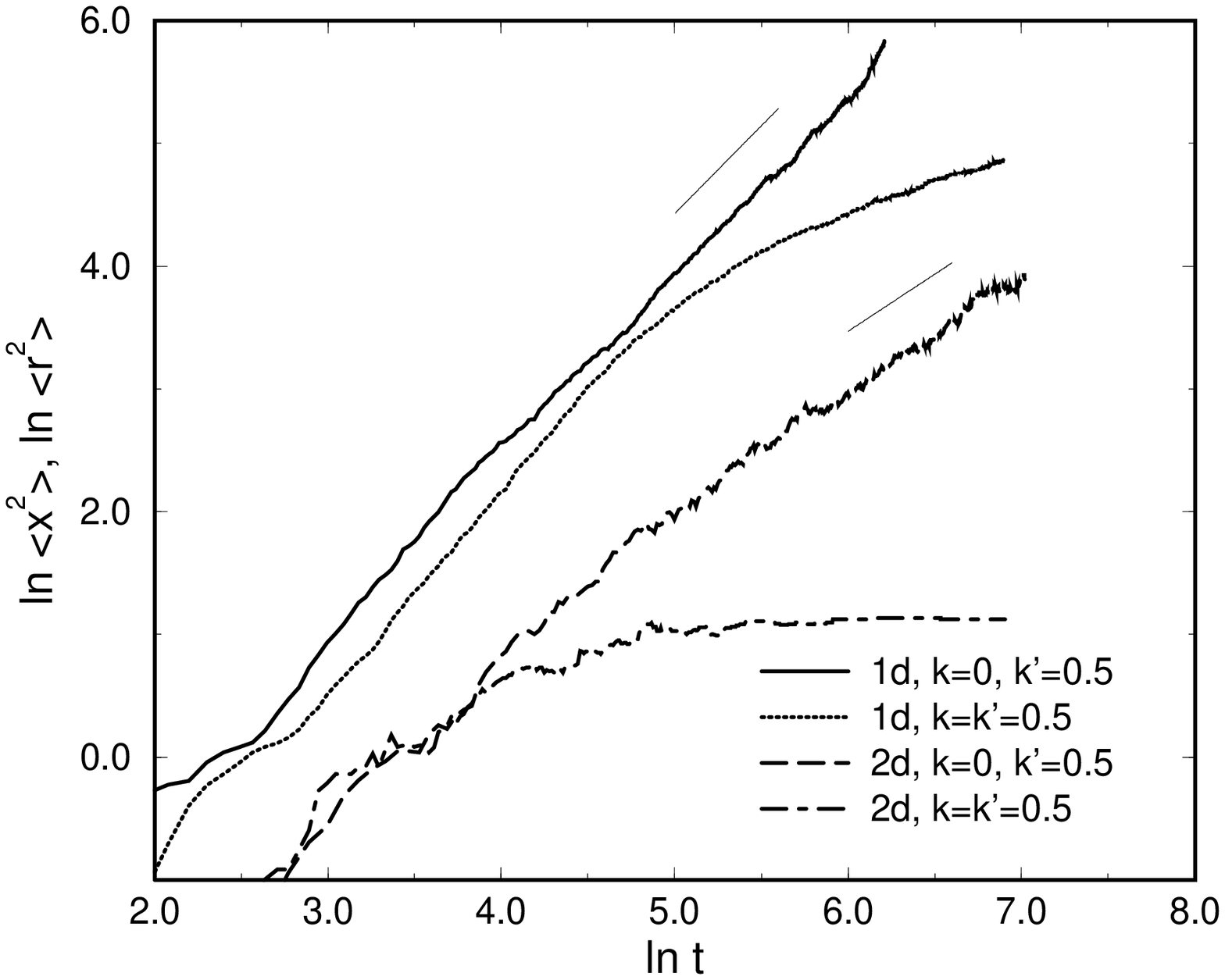}
\end{center}
\caption{Mean squared displacement of a localized mode in various
one dimensional ($300$ sites) and two dimensional ($30\times 30$) FPU
arrays.  
The slopes of the two short straight lines are $1.5$ and $0.89$. 
In all cases $A=2$, $T=0.1$, and $\gamma=0.1$.
}
\label{figrr}
\end{figure}

A typical mean squared displacement for the one-dimensional purely anharmonic
array is shown in Fig.~\ref{figrr}.  
The mean square displacement follows the superdiffusive law $\left<
x^2(t)\right> \sim t^{\alpha}$ with $\alpha=1.5$ over the entire lifetime
of the excitation.  This particular exponent is recovered for the
purely quartic chain under all conditions that we have tested, that is,
independently of force constant, excitation amplitude, and temperature.
Parameter variations affect only the prefactor, which reflects the breather
velocity.  Indeed, it does not matter {\em when} in the course of the
relaxation process the localized excitation is introduced: its mean squared
displacement grows with the same exponent $1.5$ until the excitation is
extinguished into the background.  This confirms the role of the persistent
low-frequency excitations.  We have argued~\cite{new} that this particular
superdiffusive exponent can be understood in terms of scattering events
that occur with a time distribution $\nu(t)\sim t^{-5/2}$~\cite{barkai}.

In a purely anharmonic two-dimensional array the motion is rather different
(which is consistent with the spectral differences in one and two
dimensions).  A typical mean squared displacement is also shown 
in Fig.~\ref{figrr}.  The law is now sub diffusive, 
$\left<r^2(t)\right> \sim t^{\alpha}$ with $\alpha=0.89$ over the lifetime
of the excitation.  Again, this exponent is insensitive to force constant,
excitation amplitude, and temperature changes.  It does reflect the fact
that the excitation moves much less in two dimensions than in one (although
it does of course move and eventually recedes into the background).  In 
earlier work~\cite{wepulse} we noted that a pulse in a one dimensional
purely hard array tends to move more rapidly but remains more tightly
concentrated than a pulse in, say, a harmonic or soft array.  We also noted
that in two (or more) dimensions these two tendencies are
in some sense contradictory
since the only way that a symmetric excitation can move is by breaking its
symmetry and/or dispersing.  The sort of perturbation that would set a
breather in motion requires an asymmetry that is more difficult to achieve
in two dimensions than in one.  If the distribution
of collision times of energetic breathers with other excitations that can
set it in motion has sufficiently long quiescent periods (long-tailed
waiting time probability distribution function) then the motion of the
excitations is typically subdiffusive~\cite{metzler}. Note that this does
not preclude collisions that lead to energy loss by the breather even if it
is not set in motion.

The situation in mixed anharmonic arrays is similar in one and two
dimensions, see Fig.~\ref{figrr}.  The injected excitations at first
move with the same characteristic exponents $\alpha$ as in
the corresponding purely anharmonic systems, but as the
harmonic interactions sweep the background thermal energy out of the
system,
the localized excitations stop moving.  The mean squared displacement
in each dimension then becomes independent of time
(earlier in two dimensions than in one).

\section{Summary}
\label{summary}
Energy relaxation in one- and two-dimensional nonlinear arrays with quartic
interparticle interactions (Fermi-Pasta-Ulam or FPU arrays) proceeds along
energetic pathways completely different from those of harmonic systems and
is quite sensitive to the presence or absence of quadratic contributions to
the interactions.  Relaxation in a purely harmonic array involves the
sequential decay of independent phonon modes starting with those of lowest
frequency and moving upward across the spectrum.  The decay of
energy in these arrays is exponential at short and long times, but follows
an inverse power law at intermediate times.  Throughout the decay process,
the energy is distributed uniformly over the entire array.
A localized excitation injected in the lattice simply decays according
to the distribution of characteristic relaxation times of its
phonon components.  

FPU arrays with quadratic and quartic interactions contain phonon-like
modes as well as high-energy nonlinear, and to varying degrees
localized, excitations (provided the initial temperature is sufficiently
high to excite these).  The relaxation process involves the decay of
phonons in the same spectral order as in the harmonic arrays and also
energy losses through collisions of mobile high-energy localized
excitations as they collide with lower-frequency ones.  Eventually the
harmonic interactions succeed in ``sweeping" the system clean of low energy
excitations and the remaining localized modes are quasistationary breather
solutions that persist for a very long time.  The decay to this
quasistationary state is a stretched exponential with an exponent that
depends on temperature but not on system size or damping coefficient. 
An explicitly injected high-energy localized breather follows
this behavior, with a mean squared displacement that at first grows with
time but then becomes independent of time when the system has been swept
clean of background excitations.

The greatest differences between one- and two-dimensional FPU arrays occur
in the purely quartic arrays.  In all cases the relaxation begins from the
high-frequency end of the spectrum (opposite to the harmonic case) and
involves not a direct absorption by the cold reservoir but rather a
degradation of high-frequency excitations to lower-frequency ones.  In one
dimension this degradation process is considerably faster than in
two dimensions.  The lowest frequency modes in one dimension are absorbed
by the cold reservoir but are quickly replenished by the degradation
process.  The energy relaxation is essentially exponential, with a time
constant determined by the decay of the lowest frequency components.  In
two dimensions the degradation process is slower and there are frequency
bottlenecks so that the decay of the lowest frequency excitations into the
cold reservoir no longer constitutes the rate limiting process.  Instead,
the degradation and decay contribute to a resulting stretched exponential
energy decay with an exponent that depends on system size and damping
coefficient but is independent of temperature.  In both one and two
dimensions there
remains a thermal residue of localized low-frequency excitations that
continue to perturb and degrade higher frequency ones.  The array is never
``swept clean" of low-energy excitations
as is the mixed array, and therefore no persistent breathers
occur in this system.  In order to confirm this behavior we have followed
the dynamics of an injected high-frequency localized excitation in these
arrays.  In one dimension this localized excitation remains localized but
is very mobile throughout its lifetime, being characterized by a
super-diffusive mean squared displacement.  In two dimensions the
excitation also remains localized and is much less mobile (but
nevertheless, always mobile until it disappears), being characterized by 
sub-diffusive motion.

\section*{Acknowledgments}
This work was supported in part by the
Engineering Research Program of the Office of Basic Energy Sciences at
the U. S. Department of Energy under Grant
No. DE-FG03-86ER13606. Partial support was provided by a grant from the
University of California Institute for M\'exico and the United States (UC
MEXUS) and the Consejo Nacional de Ciencia y Tecnolog\'{\i}a de
M\'exico (CONACYT), and by IGPP under project Los Alamos/DOE 822AR.

\end{document}